\documentclass[twocolumn,showpacs,preprintnumbers,amsmath,amssymb,superscriptaddress]{revtex4}

\usepackage{graphicx}
\usepackage{dcolumn}
\usepackage{bm}

\begin{document}

\preprint{Chinese Phys. B \textbf{18}, 1674 (2009).}

\title{Topological aspect of disclinations in two-dimensional melting}

\author{Wei-Kai Qi}
\affiliation{Institute of Theoretical Physics, Lanzhou University,
Lanzhou $730000$, China}

\author{Tao Zhu}
\affiliation{Institute of Theoretical Physics, Lanzhou University,
Lanzhou $730000$, China}

\author{Yong Chen}
\altaffiliation{Author to whom correspondence should be addressed.
Email: ychen@lzu.edu.cn}
\affiliation{Institute of Theoretical Physics, Lanzhou University, Lanzhou $730000$, China}
\affiliation{Key Laboratory for Magnetism and Magnetic materials of the Ministry of Education, Lanzhou University, Lanzhou $730000$, China}

\author{Ji-rong Ren}
\affiliation{Institute of Theoretical Physics, Lanzhou University,
Lanzhou $730000$, China}

\date{\today}

\begin{abstract}
By using topological current theory, we study the inner
topological structure of disclinations during the melting of two-dimensional systems. From two-dimensional elasticity theory, it is found topological currents for topological defects in homogeneous
equation. The evolution of disclinations is studied, and the branch
conditions for generating, annihilating, crossing, splitting and
merging of disclinations are given.
\end{abstract}

% 64.70.D-, SolidCliquid transitions
% 82.70.Dd, colloids
% 61.72.Lk, Linear defects: dislocations, disclinations

\maketitle

%\section{\label{sec:level1}INTRODUCTION}
Topological defects, which are a necessary consequence of broken
continuous symmetry, play an important role in two-dimensional phase
transition. In 1970's, Kosterlitz and Thouless construct a detailed
and complete theory of superfluidity on two-dimensions~\cite{kt}.
They indicate vortices pair unbinding will lead to a second-order
transition in superfluid films. Later, a microscopic scenario of 2D melting has been posited in the
form of the Kosterlitz-Thouless-Halperin-Nelson-Young (KTHNY)
theory~\cite{two,nh,Yp}. The KTHNY theory predicts a new phase, the
so-called hexatic phase,  that exists between the solid and liquid
phases in 2D melting~\cite{Dn}.

In two-dimensional colloid systems, topological defects have been studied
 in experiments and computer simulations. A serial experiments were
performed to calculate dislocations and disclinations dynamic of
two-dimensional colloidal systems, and dissociation of dislocations and disclinations were observed~\cite{My,Ta,Mar,soft}. During the years, a
large number of computer simulations indicated that exist a two-stage melting
as prescribed by KTHNY theory, however, results are still controversial~\cite{ja,bo,Kn,Ns}. Although the KTHNY theory is currently preferred, a different theoretical approach, evoking grain-boundary-induced melting, was a
first-order transition suggested by Chui~\cite{chui}. One may note that the
condensation of geometrical defects is also a first-order
transition~\cite{Gd1,Gd2}. Our previous work found that exist a hexatic-isotropic liquid phase coexistence during the melting of soft Yukawa systems~\cite{Qi,XQ}. By Voronoi polygons analysis, the behavior of piont defects in the coexistence is very complicated. The evolution of topological defects during the melting of two-dimensional system still a open question.

Recently, a topological field theory for topological defects developed
by Duan et al\cite{Duan01}. By using
Duan's topological current theory, the inner topological
structure and bifurcation of topological defects, such as
disclination and dislocation in liquid crystal and soild, were
studied. In KT phase transition, there also exists an
elementary vortex topological current constructed by the superfluid
order parameter\cite{Duan02}. By using the topological current theory, we can give the the branch conditions for generating, annihilating, crossing, splitting and
merging of topological defects.

In this paper, we will discuss the topological quantization and
bifurcation of topological defects in two-dimensional crystals. This
work is based on the so-called Duan's
topological current theory. The organization of this paper is as follows. We  describe the elasticity theory. Using Duan's
topological current theory, we discussed the topological structure of disclination in two-dimensional crystals. In the last section, we summarize our results.

%\section{\label{sec:level1}Two-dimensional elasticity theory}
In continuum elasticity theory, the elastic Hamiltonian in
two-dimensional triangular solid is given by~\cite{Landau}
\begin{equation}
F=\frac{1}{2}\int d^2 r(2\mu u_{ij}^2+\lambda u_{kk}^2),
\end{equation}
where $\lambda$ and $\mu$ are the two-dimensional Lam\'e
coefficients. The strain tensor is
\begin{equation}
u_{ij}(r)=\frac{1}{2}\bigg[\frac{\partial u_i(r)}{\partial
r_j}+\frac{\partial u_j(r)}{\partial r_i  }\bigg]
\end{equation}
 A deformation is represented by a displacement vector field
$\textbf{u}(r)=(u_1,u_2)$, which maps the point $\textbf{r}=(x,y)$ to
$\textbf{r}+\textbf{u}$. If there are no defects, the deformation is
a single-valued mapping of the plane onto itself. But \textbf{u}
becomes a multi-valued function when there is a dislocation. A single
dislocation corresponds to an extra half lattice plane, which
characterized by a Burger's vector \textbf{b}. Another type of defect
in two-dimension crystal is disclination, which is defined in terms
of the bond angle field $\theta$. $\theta(x)$ is the angle between
local lattice bonds and a reference axis.

If we minimize $F_s$ with respect to variations in \textbf{u}, we
obtain the equation
\begin{equation}
\partial_i\sigma_{ij}=0,
\label{eq:sigma}
\end{equation}
Where the stress tensor $\sigma_{ij}$ is defined by
\begin{equation}
\sigma_{ij}=2\mu u_{ij}+\lambda u_{kk}\delta_{ij},
\end{equation}
Because $\sigma_{ij}$ is symmetric, it can be written as
\begin{equation}
\sigma_{ij}=\epsilon_{ik}\epsilon_{jl}\nabla_k\nabla_l\chi,
\end{equation}
The function $\chi$ is called the Airy stress function. Although any
choice for $\chi$ yields a stress tensor that satisfies
Eq.(\ref{eq:sigma}), the choice cannot be arbitrary. The strain
$u_{ij}$ is related to the stress is
\begin{eqnarray}
u_{ij}&=&\frac{1}{2\mu}\sigma_{ij}-\frac{\lambda}{4(\lambda+\mu)}\delta_{ij}\sigma_{kk}
\nonumber\\
&&=\frac{1+\sigma_{0}}{Y}\epsilon_{ik}\epsilon_{kl}\nabla_{k}\nabla_{l}\chi-\frac{\sigma_{0}}{Y}\nabla^{2}\chi\delta_{ij}
\end{eqnarray}
where $Y=4\mu(\mu+\lambda)/(2\mu+\lambda)$ is the two-dimensional
Young's modulus and $\sigma_{0}=\lambda/(2\mu+\lambda)$ the
two-dimensional Poisson ratio. Applying
$\epsilon_{ik}\epsilon_{jl}\nabla_k\nabla_l$ to both of this
equation, we find
\begin{eqnarray}
&\frac{1}{Y}\nabla^4\chi&=\frac{1}{2}\epsilon_{ik}\epsilon_{jl}\nabla_k\nabla_l(\partial_i
u_j+\partial_j u_i)\nonumber\\
&&=\frac{1}{2}\epsilon_{ik}\epsilon_{jl}\nabla_k\nabla_l(\partial_i
u_j-\partial_j u_i)
+\frac{1}{2}\epsilon_{ik}\epsilon_{jl}\nabla_k\nabla_l\partial_j
u_i\nonumber\\
&&=\epsilon_{kl}\partial_{k}\partial_{l}\theta+\epsilon_{ik}\partial_{k}(\epsilon_{jl}\partial_l\partial_j
u_i) \label{eq:77}
\end{eqnarray}

%\section{\label{sec:level2}Topological Current in two-dimensional crystal}
The defects associated with the continuum elastic theory of a solid
are dislocations and disclinations. Dislocations and disclinations
can be introduced into the theory in a way similar to the discussion
of superfluid vortices\cite{Duan04}. In the following we consider
only the triangular lattice since it is the most densely packed one
in two-dimensional and favored by Nature.

Disclinations, which are characterized by a topological charge, have
a much higher energy than dislocations. They are defined in terms of
the bond angle field $\theta(r)$, which measures the bonds
orientation. It is convenient to define an order parameter for bond
orientations, which for the triangular lattices is
$\psi(r)=\psi_0e^{i6\theta(r)}$. However, the bond angle field is
undefined at the disclination cores, i.e., the zero points of the
order parameter. We rewrite the orientation order parameter
$\psi(r)=\phi_6^1+i\phi_6^2$ instead of
$\psi(r)=\psi_0e^{i6\theta(r)}$. Let us define the unit
vector field $\vec{n}$ as
\begin{equation}
n^a=\frac{\phi_6^a}{||\phi_6||},~||\phi_6||=\sqrt{\phi_6^a\phi_6^a},~a=1,2.
\end{equation}
Obviously, $n^a n^a=1$. The topological defect is related to the
zero points of the two-component vector parameter $\Psi$, i.e.,
\begin{eqnarray}
\phi_6^1(x,y)=0,~\phi_6^2(x,y)=0. \label{eq:zero}
\end{eqnarray}

Suppose there is a defect located at $z_i$, the topological charge of
the defect is defined by the Gauss map n: $\partial\sum_i\rightarrow
S^1$,
\begin{equation}
W(\phi_6,z_i)=\frac{1}{2\pi}\oint_{\partial\sum_i}\epsilon_{ab}n^a dn^b
\end{equation}
Using the stokes' theorem in the exterior differential form, one can
deduce that
\begin{equation}
W(\phi_6,z_i)=\frac{1}{2\pi}\oint_{\sum_i}\epsilon_{ab}\epsilon^{ij}\partial_{i}n^a
\partial_{j}n^b d^2x
\end{equation}

We can deduce a topological current of disclinations in two-dimensional crystal,
\begin{equation}
j_{disc}^k=\frac{1}{6}\epsilon^{ijk}\epsilon_{ab}\partial_i n_6^a\partial_j
n_6^b=\delta^2(\vec{\phi_6})J^k\big(\frac{\phi_6}{x}\big)
\end{equation}
It is the $\phi$-mapping current for disclination. where $J^{k}(\phi_6/x$) is the vector Jacobians of $\vec{\phi_6}$,
\begin{equation}
J^{k}\big(\frac{\phi_6}{x}\big)=\frac{1}{2}\epsilon^{ijk}\epsilon_{ab}\partial_{i}\phi_6^a\partial_{j}\phi_6^b,
\label{eq:Jacobian}
\end{equation}

It is easy to see that this topological current is identically
conserved, i.e.,
\begin{equation}
\partial_{k}j^{k}=0.
\label{eq:conserved}
\end{equation}

\begin{figure}
\begin{center}
\includegraphics[width=0.15\textwidth]{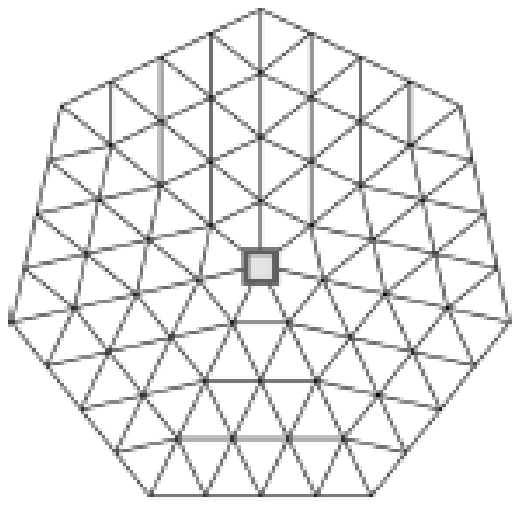}
~~\includegraphics[width=0.15\textwidth]{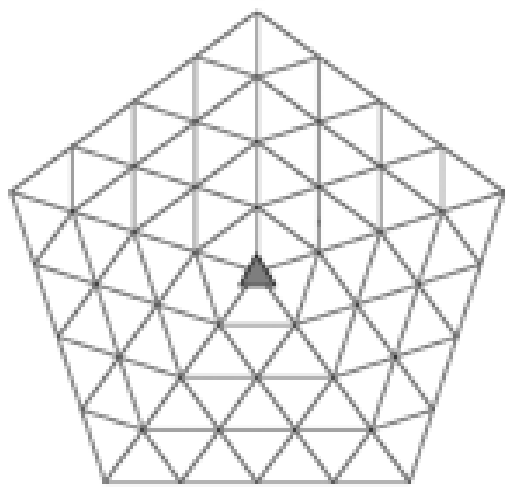}
\caption{Disclination in triangular lattices. A $-\pi/3$ disclination
with its sevenfold coordinated site in the center (left). A $\pi/3$
disclination with its fivefold coordinated site (right). }
\label{Fig:03}
\end{center}
\end{figure}

According to the implicit function theorem, if Jacobian determinant
\begin{equation}
J^0\big(\frac{\phi_6}{x}\big)=J\big(\frac{\phi_6}{x}\big)\neq0,
\end{equation}
the solutions of Eq.(\ref{eq:zero}) can be generally expressed as
\begin{equation}
x=x_l(t),~y=y_l(t),~l=1,2,...,N,
\end{equation}
which represent N zero points $\vec{z_l}(t)$ (l=1,2,...,N) or world
line of N diclinations $D_l$ in space-time.
\begin{figure}
\begin{center}
\includegraphics[width=0.25\textwidth]{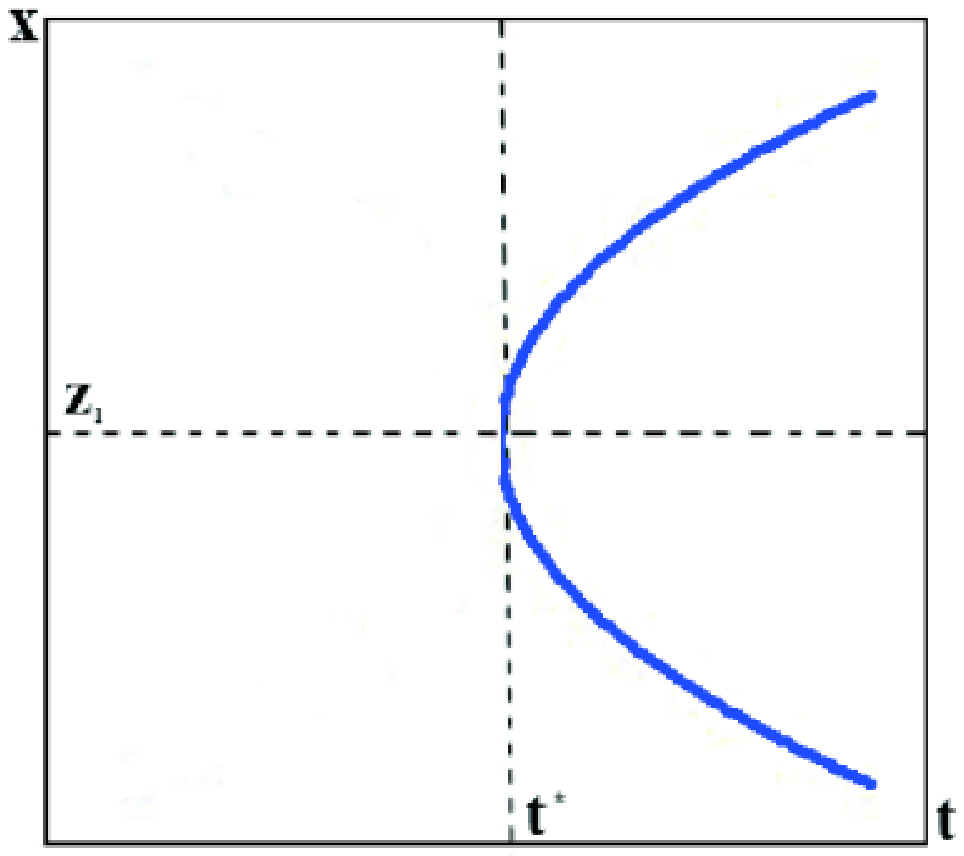}
\includegraphics[width=0.25\textwidth]{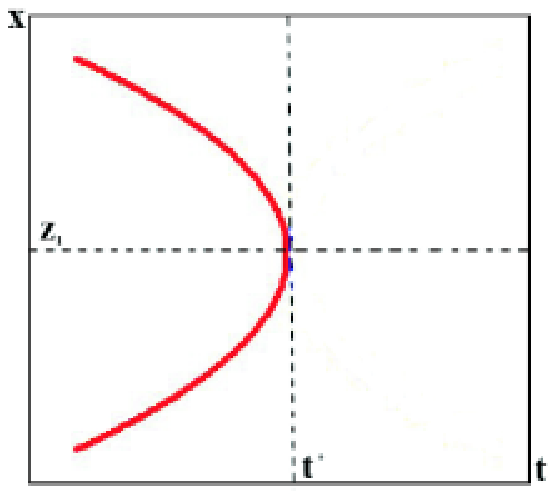}
\caption{Generating and annihilating of disclination pairs. }
\label{Fig:03}
\end{center}
\end{figure}

With the $\delta$-function theory, $\delta^2(\phi)$ can be expanded
as
\begin{equation}
\delta^2(\phi_6)=\sum_{l=1}^N
\frac{\beta_l}{|J(\phi_6/x)_{z_l}|}\delta^2(\vec{r}-\vec{z_l}(t))
\end{equation}
where the positive integer $\beta_l$ is called the Hopf index of map
$x\rightarrow\vec{\phi}$. The meaning of $\beta_l$ is that when the
point $\vec{r}$ covers the neighborhood of the zero $\vec{z_l}$ once,
the vector field $\vec{\phi_6}$ covers the corresponding region for
$\beta_l$ times. Using the implicit function theorem and the
definition of vector Jacobians (\ref{eq:Jacobian}), we can get the
velocity of the l-th defect,
\begin{eqnarray}
\vec{v_l}=\frac{d\vec{z_l}}{dt}=\big[\frac{\vec{J}(\phi_6/x)}{J(\phi_6/x)}\big]_{\vec{z_l}}
\nonumber\\
\vec{J}\big(\frac{\phi_6}{x}\big)=\big[J^1\big(\phi_6/x\big),
J^2\big(\phi_6/x\big)\big]
\end{eqnarray}
Then the spatial and temporal components of the defect current $j^u$,
can be written as the form of the current and the density of the
system of N classical points particles with topological charge $W_l=\beta_l\eta_l$ moving in the
(2+1)-dimensional space-time,
\begin{equation}
\vec{j}=\sum_{l=1}^{N}\beta_l\eta_l\vec{v_l}\delta^2(\vec{r}-\vec{z_l}(t))
\label{eq:current}
\end{equation}
\begin{equation}
\rho=\sum_{l=1}^{N}\beta_l\eta_l\delta^2(\vec{r}-\vec{z_l}(t))
\end{equation}
where $\eta_l$ is Brouwer degree,
\begin{equation}
\eta_l=\frac{J(\phi_6/x)}{|J(\phi_6/x)|}\bigg|_{\vec{z_l}}=\pm1
\end{equation}

For disclinations, using Duan's topological current theory the
homogeneous equation can write as
\begin{equation}
\frac{1}{Y}\nabla^4\chi=\frac{2\pi}{6}\delta^2(\vec{\phi_6})J\big(\frac{\phi_6}{x}\big)
\end{equation}
Similar results get by Nelson in the KTHNY theory\cite{Nelson}. In our theory, the
topological charge of a disclination $D_l$ is
\begin{eqnarray}
Q_l&=&\frac{1}{6}\oint_{\Sigma_i}\epsilon_{ab}\epsilon^{ij}\partial_in_6^a\partial_jn_6^bd^2x\nonumber
\\&&=\frac{2\pi}{6}W_l=\beta_l\eta_l
\label{tc}
\end{eqnarray}
where $W_l$ is the winding number of $\Psi$ around $D_l$, the above expression reveals distinctly that the topological charge of disclination is not only the winding number, but also expressed by the Hopf indices and Brouwer degrees. The topological inner structure showed in Eq.(\ref{tc}) is more essential than that
in Eq.(\ref{eq:77}), this is the advantage of our topological description of the disclination.

It is clearly seen that Eq.(\ref{eq:current}) shows the movement of
two-dimension crystal topological defects in space-time. According to
Eq.(\ref{eq:conserved}), the topological charge of defects in
two-dimensional crystal are conserved,
\begin{equation}
\frac{\partial \rho}{\partial t}+\nabla\vec{j}=0.
\end{equation}
In addition, there is a constraint of "charge neutrality",
\begin{equation}
\int\rho d^2x=\frac{1}{2\pi}\sum_{l=1}^{N}\beta_{l}\eta_{l}=0.
\end{equation}
It indicate that the defect in two-dimensional crystal appear in
pair.

\begin{figure}
\begin{center}
\includegraphics[width=0.25\textwidth]{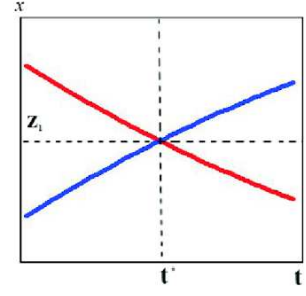}
\caption{Two disclinations collide with different directions of
motion at the bifurcation point in (2+1)-dimensional space-time. }
\label{Fig:04}
\end{center}
\end{figure}

Analogy to vortices in superfluid films, the zero point of the order
parameter field play an important role in describing the topological
defects in two-dimensional crystal. Now we study the properties of
these zero points. If the Jacobian determinant $J^0(\phi_6/x)\neq0$,
we will have the isolated solution ofthe zeros of the order
parameter field. But when $J^0(\phi_6/x)=0$, the above results will
change in some way, and will lead to the branch process of defects.
We denote one of the vectors Jacobian at zero points as $(t^*,
\vec{z_l})$. According to the values of the vector Jacobian at zero
points of the order parameter, there are limit points and
bifurcation points. Each kind corresponds to different cases of
branch processes.

\begin{figure}
\begin{center}
\includegraphics[width=0.25\textwidth]{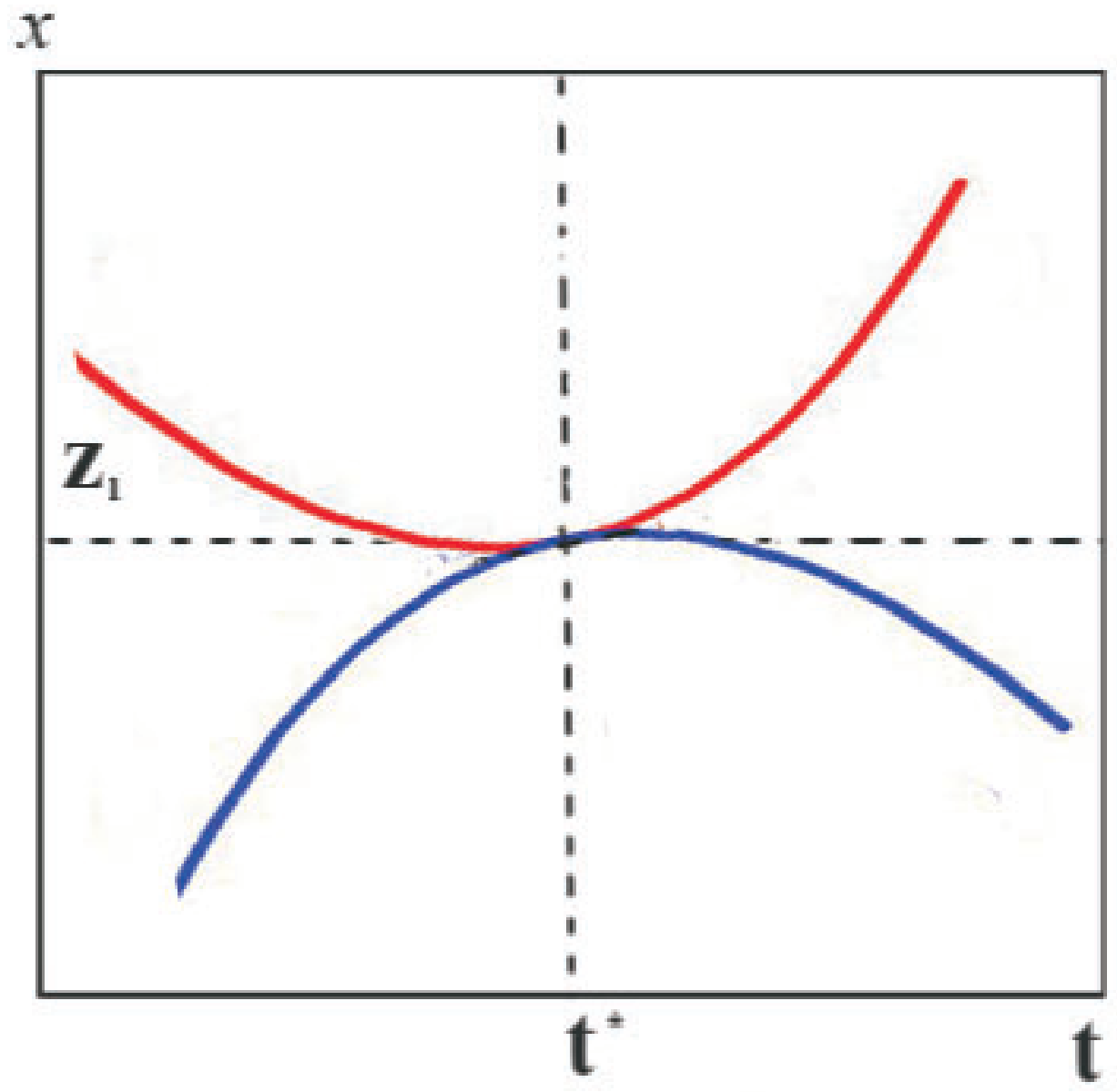}
\includegraphics[width=0.25\textwidth]{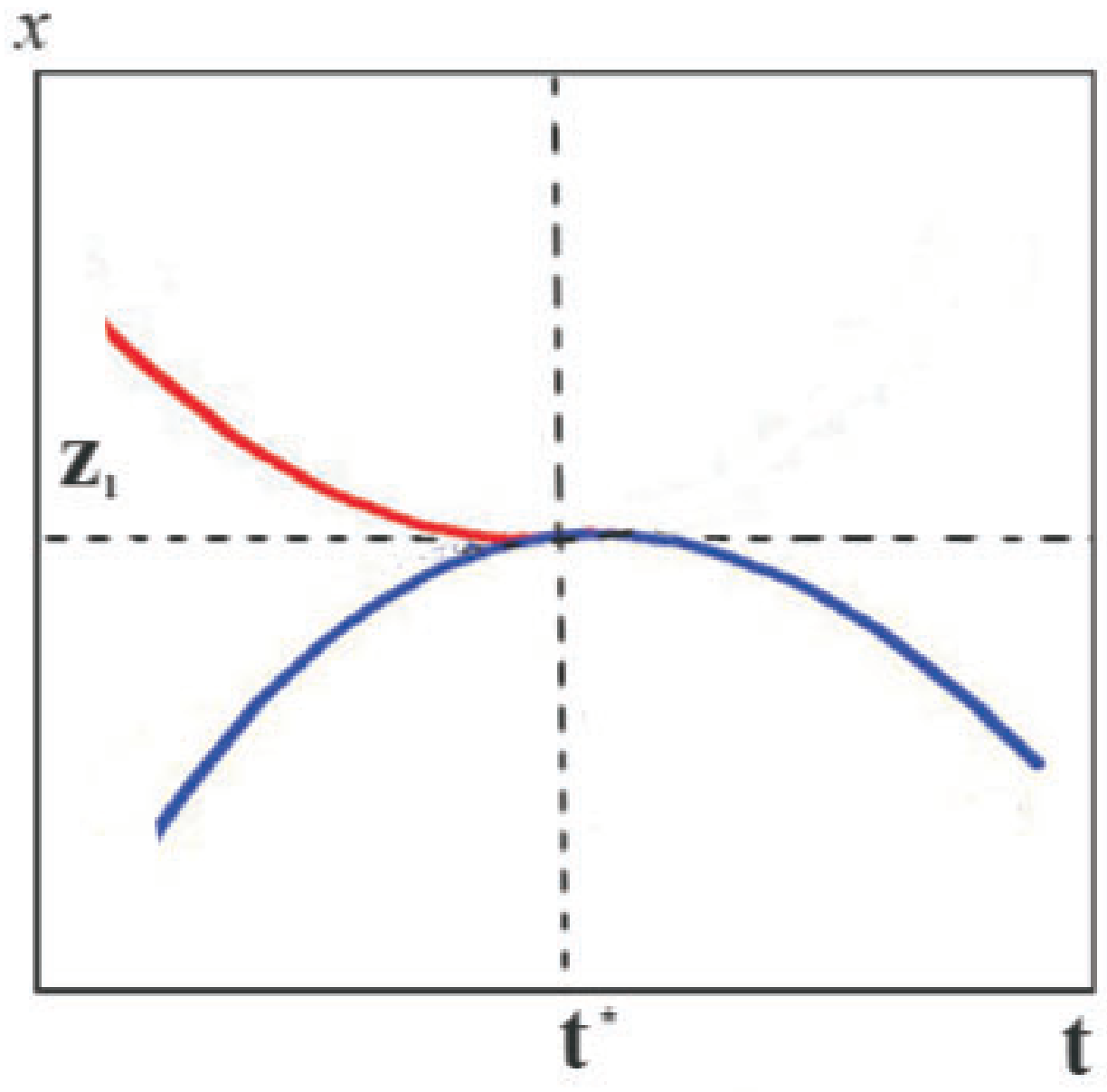}
\includegraphics[width=0.25\textwidth]{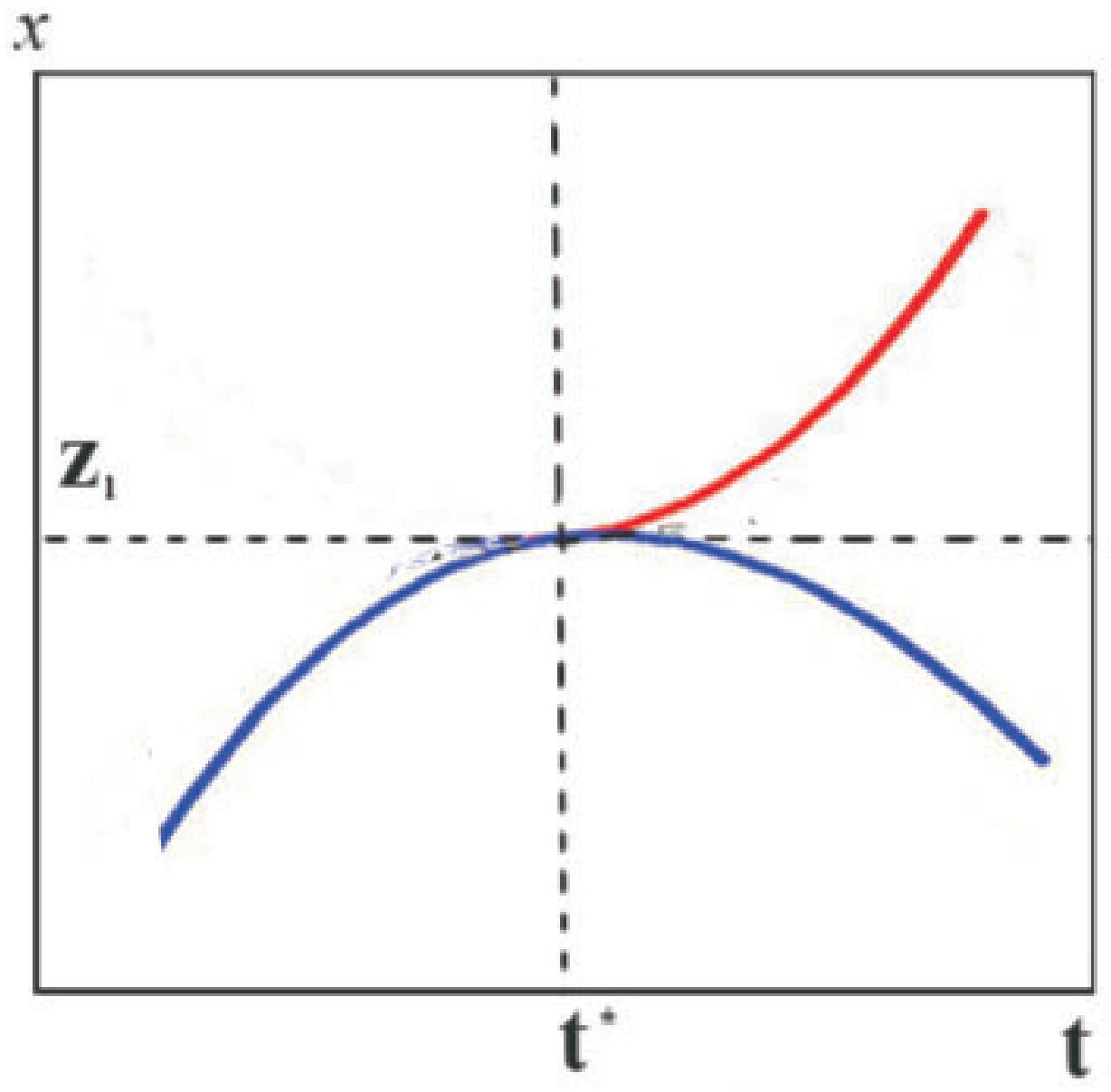}
\caption{Two disclinations collide with different directions of
motion at the bifurcation point in (2+1)-dimensional space-time. }
\label{Fig:04}
\end{center}
\end{figure}

Let us explore what will happen to the disclination at the limit
point $(t^*, \vec{z_l})$. The limit points are determined by
\begin{equation}
J^{0}\big(\frac{\phi_6}{x}\big)\big|_{t^*, \vec{z_l}}=0,~
J^{1}\big(\frac{\phi_6}{x}\big)\big|_{t^*, \vec{z_l}}\neq0
\label{eq:condition01}
\end{equation}
\begin{equation}
J^{0}\big(\frac{\phi_6}{x}\big)\big|_{t^*, \vec{z_l}}=0,~
J^{2}\big(\frac{\phi_6}{x}\big)\big|_{t^*, \vec{z_l}}\neq0
\end{equation}
Considering the condition (\ref{eq:condition01}) and making use of
the implicit function theorem, the solution of Eq.(\ref{eq:zero}) in
the neighborhood of the point ($t^*, \vec{z_l}$),
\begin{equation}
t=t(x),~y=y(x)
\end{equation}
where $t^*=t(z_l^1)$. In this case, one can see that
\begin{equation}
\frac{dx}{dt}\bigg|_{(t^*,
\vec{z_l})}=\frac{J^1(\phi_6/x)}{J(\phi_6/x)}\bigg|_{(t^*,
\vec{z_l})}=\infty,
\end{equation}
or
\begin{equation}
\frac{dt}{dx}\bigg|_{(t^*, \vec{z_l}}=0.
\end{equation}
The Taylor expansion of $t=t(x)$ at the limit points $(t^*,
\vec{z_l})$ is
\begin{equation}
t-t^*=\frac{1}{2}\frac{d^2t}{dx^2}\bigg|_{t^*, \vec{z_l}}(x-z_l^1)^2,
\end{equation}
which is a parabola in the x-t plane. From this equation, we can
obtain two solutions $x_1(t)$ and $x_2(t)$, which give two branch
solutions (World lines of disclinations). If
$$
\frac{d^2t}{dx^2}\bigg|_{(t^*, \vec{z_l})}>0,
$$
we have the branch solutions for $t > t^*$. It is related to the
origin of a disclination pair. Otherwise, we have the branch
solutions for $t < t^*$, which related to the annihilation of a
disclination pair.

Since the topological current is identically conserved, the
topological charges of these two generated or annihilated
disclinations must be opposite at the limit points, say
\begin{equation}
\beta_1\eta_1+\beta_2\eta_2=0.
\end{equation}
For a limit point it is required that $J^1(\phi/x)|_{t^*,
\vec{z_l}}\neq0$. As to bifurcation point, it must satisfy a more
complex condition. This case will be discussed in the following.

%\subsection{Encountering, Splitting and Merging of Disclinations}
Now, let us turn to consider in which the restrictions on zero point
$(t^*, \vec{z_l})$ are
\begin{equation}
J^{k}\bigg(\frac{\phi_6}{x}\bigg)\bigg|_{(t^*, \vec{z_l})}=0,
~~k=0,1,2,\label{eq:con}
\end{equation}
which imply an important fact that the function relationship between
t and x or y is not unique in the neighborhood of the bifurcation
point $(t^*, \vec{z_l})$. This fact is easily seen from
\begin{equation}
\frac{dx}{dt}=\frac{J^1(\phi_6/x)}{J(\phi_6/x)}\bigg|_{t^*, \vec{z_l}},~~
\frac{dy}{dt}=\frac{J^2(\phi_6/x)}{J(\phi_6/x)}\bigg|_{t^*,
\vec{z_l}},\label{eq:con2}
\end{equation}
which under Eq.(\ref{eq:con}) directly shows the indefiniteness of
the direction of integral curve of Eq.(\ref{eq:con2}) at $(t^*,
\vec{z_l})$. This is why the very point $(t^*, \vec{z_l})$ is called
a bifurcation point of the orientation order parameter.

With the aim of finding the different directions of all branch curves
at the bifurcation point, we suppose
\begin{equation}
\frac{\partial \phi_6^1}{\partial y}\bigg|_{t^*, \vec{z_l}}\neq0.
\end{equation}
According to the $\phi$-mapping theory, the Taylor expansion of the
solution of the zeros of the order
parameter field in the neighborhood of $(t^*,
\vec{z_l})$ can be expressed as
\begin{equation}
A(x-z_l^1)^2+2B(x-z_l^1)(t-t^*)+C(t-t^*)^2+...=0,
\end{equation}
which leads to
\begin{equation}
A\big(\frac{dt}{dx}\big)^2+2B\frac{dx}{dt}+C=0, \label{eq:two01}
\end{equation}
and
\begin{equation}
C\big(\frac{dx}{dt}\big)^2+2B\frac{dt}{dx}+A=0,
 \label{eq:two02}
\end{equation}
where A, B, and C are constants determined by the order parameter. The solutions of Eq.(\ref{eq:two01}) or
Eq.(\ref{eq:two02}) give different directions of the branch curves
(world line of vortices) at the bifurcation point. There are four
possible cases, which will show the physical meanings of the
bifurcation points.

\begin{figure}
\begin{center}
\includegraphics[width=0.25\textwidth]{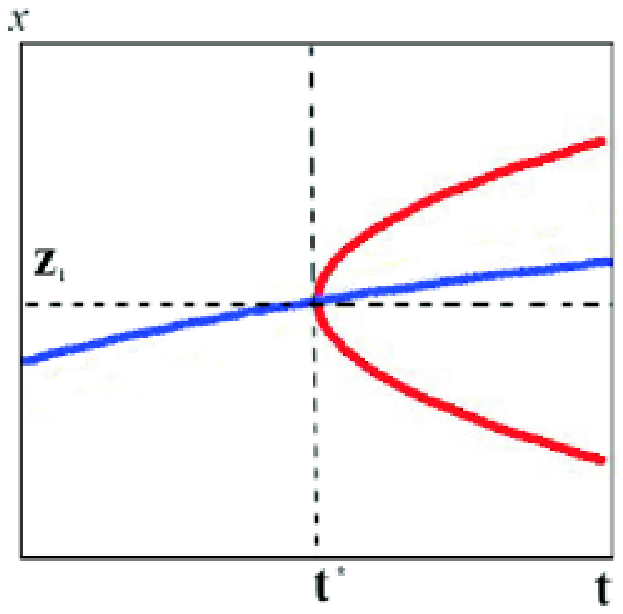}
\includegraphics[width=0.25\textwidth]{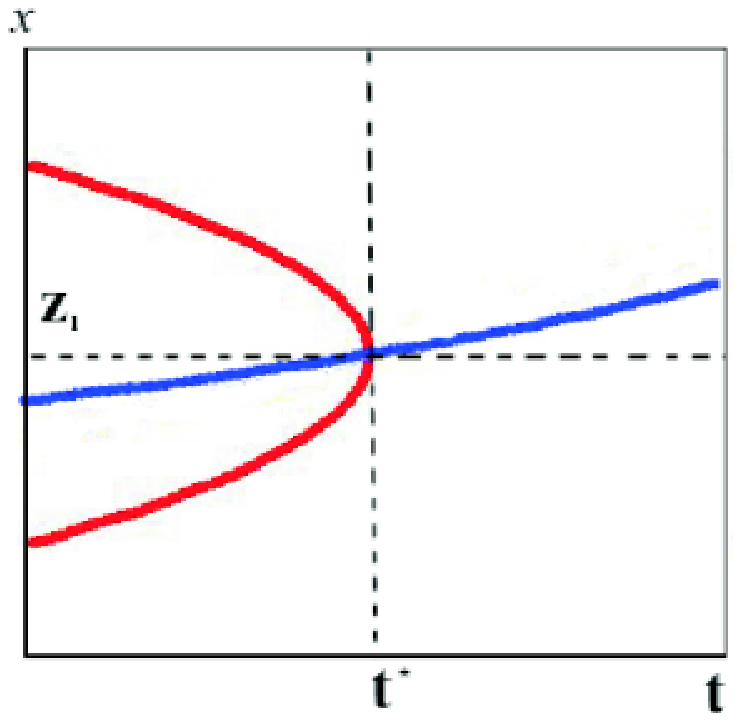}
\caption{Two disclinations collide with different directions of
motion at the bifurcation point in (2+1)-dimensional space-time. }
\label{Fig:04}
\end{center}
\end{figure}

Case 1 $(A\neq0)$. For $\Delta=4(B^2-AC)>0$, from Eq.(\ref{eq:two01})
we get two different motion directions of the core of disclination
\begin{equation}
\frac{dx}{dt}\big|_{1,2}=\frac{-B\pm\sqrt{B^2-AC}}{A},
\end{equation}
where two world lines of two
disclination intersect which different directions at the bifurcation
point. This shows that two disclinations encounter and the depart at
the bifurcation point.

Case 2 $(A\neq0)$. For $\Delta=4(B^2-AC)>0$, form
Eq.(\ref{eq:two01}), we obtain only one motion direction of the core
of disclination
\begin{equation}
\frac{dx}{dt}\bigg|_{1,2}=-\frac{B}{A},
\end{equation}
which includes three important cases. (i) Two world lines
tangentially contact, i.e., two disclinations tangentially encounter
at the bifurcation point. (ii) Two world lines merge into one world
line, i.e., two disclinations merge into one disclination at the
bifurcation point. (iii) One world line resolves into two world
lines, i.e., one disclinations splits into two disclinations at the
bifurcation point.

Case 3 $(A=0, C\neq=0)$ For $\Delta=4(B^2-AC)=0$ from
Eq.(\ref{eq:two01}) we have
\begin{equation}
\frac{dt}{dx}\bigg|_{1,2}=\frac{-B\pm\sqrt{B^2-AC}}{C}=0,
~-\frac{2B}{C}.
\end{equation}
There are two important cases: (i) One world line resolves into three
world lines, i.e., one disclination split into three disclinations at
the bifurcation point. (ii) Three world line merge into one world
line, i.e., three disclinations merge into one disclination at the
bifurcation point.

Case 4 (A=C=0). Equations (\ref{eq:two01}) and (\ref{eq:two02}) give
respectively
\begin{equation}
\frac{dx}{dt}=0,~\frac{dt}{dx}=0.
\label{eq:case4}
\end{equation}
This case shows that two worldlines intersect normally at the
bifurcation point, which is similar to case 3. It is no surprise that
both parts of Eq.(\ref{eq:case4}) are correct because they give the
slope coefficients of two different curves at the same point $(t^*,
\vec{z_l})$.

 In conclusion, we study the inner topological structure of disclinations in
two-dimensional colloid systems. We have obtained a more essential topological formulary of charge density of disclinations in two-dimensional crystals, and revealed the  inner topological relationship of the charge of disclinations is characterized by the Hopf index and the Brouwer degree. We have studied the evolution of disclinations by making use of Duan's topological current theory. We concluded that there exist crucial cases of branch processes in the evolution of disclinations when $J(\phi/x)=0$, i.e, $\eta_l$ is indefinite. It means that disclinations are generated or annihilated at the limit points and are encountered, split, or merge at the bifurcation points, which shows that the disclination is unstable at these branch points. We would like to pointed that all the results in this paper obtains from the viewpoint of topology without any hypothesis, and they are not depended on the property of systems, such as interaction between particles.

\begin{acknowledgments}
This work was supported by the SRF for ROCS, SEM, and by the Interdisciplinary Innovation Research Fund for Young Scholars,
Lanzhou University.
\end{acknowledgments}

\end{document}